\documentclass[10pt,twocolumn,english,conference]{IEEEtran}
\usepackage[T1]{fontenc}
\usepackage{array}
\usepackage{graphicx}

\makeatletter

\providecommand{\tabularnewline}{\\}

 \newtheorem{thm}{Theorem}
 \newtheorem{example}{Example}

\author{
\authorblockN{Chris Winstead}\authorblockA{ECE Dept.\\Utah State University\\4120 Old Main Hill\\Logan, UT 84322-4120\\USA\\Fax: (435) 797-3054\\Email: \textit{winstead@engineering.usu.edu}}%
\and%
\authorblockN{Vincent C. Gaudet, Anthony Rapley, Christian Schlegel}\authorblockA{ECE Dept.\\2nd Floor ECERF bldg.\\University of Alberta\\Edmonton, AB T6G 2V4\\Canada\\Email: \textit{[vgaudet,rapley,schlegel]@ece.ualberta.ca}}%
}

\usepackage{cite}

\usepackage{babel}
\makeatother
\begin{document}

\title{Stochastic Iterative Decoders}

\maketitle
\begin{abstract}
This paper presents a stochastic algorithm for iterative error control
decoding. We show that the stochastic decoding algorithm is an approximation
of the sum-product algorithm. When the code's factor graph is a tree,
as with trellises, the algorithm approaches maximum a-posteriori decoding.
We also demonstrate a stochastic approximations to the alternative
update rule known as successive relaxation. Stochastic decoders have
very simple digital implementations which have almost no RAM requirements.
We present example stochastic decoders for a trellis-based Hamming
code, and for a Block Turbo code constructed from Hamming codes.
\end{abstract}

\section{Introduction}

Iterative decoding describes a class of powerful graph-based algorithms
for error control decoding, including Turbo and LDPC decoders. Implementations
tend to be complex, and much research effort has been invested to
produce less complex physical solutions \cite{Bekooij:TurboDecoder2001,Hagenauer:ISIT1998,Kaza:TurboDecoders2004,Loeliger:ITTrans2001}. 

We describe implementation strategies in terms of \emph{parallel}
and \emph{serial} architectures. The parallel approach requires large
numbers of small processors which operate concurrently. This approach
consumes a large amount of silicon area. 

In serial architectures, processors are reused multiple times within
an iteration, thus requiring use of a RAM for storing messages. This
approach requires less silicon area, but reduces the decoder's maximum
throughput. The serial approach is also expensive to apply in reconfigurable
implementations, because RAM tends to be limited in current FPGA platforms.

In this paper, we present a stochastic approximation to the sum-product
algorithm. This approximation allows for low-complexity parallel decoder
implementations. The resulting circuits have no RAM requirements,
and are simple enough to be implemented in low-cost FPGAs.

In stochastic implementations, each processor is implemented by a
simple logic gate. The stochastic algorithm can therefore be implemented
on a code's factor graph, using only a few gates per node. 

Similar algorithms have been presented elsewhere \cite{Gaudet:Stochastic_El_Lett_2003}.
The stochastic decoding algorithm is related to Pearl's method of
{}``stochastic simulation,'' which provides accurate belief propagation
in cyclic Bayesian networks \cite{Pearl:ProbabilisticReasoning1988}.
Pearl's method does not work for cyclic networks with deterministic
constraints, and therefore cannot be used for error-control decoding.

In this paper, we present a new form of the stochastic algorithm which
is more general than the original, strictly binary algorithm \cite{Gaudet:Stochastic_El_Lett_2003}.
Unlike Pearl's method, our goal is not to produce cycle-independent
results, but to produce a low-complexity approximation of the sum-product
decoding algorithm.

This paper is organized as follows. In Sec. \ref{sec:Message-Passing-Decoding-Algorithms}
we present a review of notation and terminology. Sec. \ref{sub:The-sum-product-update}
summarizes the widely-used \emph{sum-product} iterative update rule.
Sec. \ref{sub:The-relaxation-update} presents the less common \emph{successive
relaxation} update rule, a derivative of sum-product which sometimes
yields better performance and/or faster convergence \cite{Hemait:DynamicsSOR_ISIT2004}.

In Sec. \ref{sec:The-Stochastic-Algorithm} we describe the new stochastic
decoding algorithm. Stochastic implementation on acyclic code graphs
is discussed in Sec. \ref{sub:Message-passing-implementation-on},
which also presents results for a stochastic trellis decoder. 

Cyclic constraint graphs require use of a modified node, called the
supernode, which is discussed in Sec. \ref{sub:Graphs-with-cycles}.
This Section also presents results for a length-256 stochastic Block
Turbo decoder. 

Throughout the paper, significant results are presented as theorems.
The proofs of these theorems are intended to be illustrative, but
not fully rigorous.

\section{\label{sec:Message-Passing-Decoding-Algorithms} Message-Passing
Decoding Algorithms}

The conventional sum-product algorithm consists of \emph{probability
propagation} through nodes in a code's \emph{factor graph} \cite{Kschischang:FactorGraphsIT2001}.
For our purposes, we consider a more limited class of factor graphs
called \emph{constraint graphs}. 

A constraint graph represents the relationships among a code's \emph{variables}
and \emph{constraints}. Constraint functions define which combinations
of variables are permissible and which are not.

\subsection{\label{sub:Constraint-graphs.}Constraint graphs.}

A constraint graph is a type of factor graph which shows the constraints
among a code's information and parity bits. The constraint graph consists
of nodes and edges. Variables are usually represented by circles,
and constraints (functions) are represented by squares.

It is customary to show information and parity bits explicitly as
circles in the constraint graph. These variables represent observable
information, and have a single edge connection. 

A code is constructed from a family of constraint functions, represented
in the graph by squares. For most codes of interest, all constraints
may be expressed using functions of no more than three variables. 

We now proceed to a more precise explanation of these concepts. Let
$A$, $B$ and $C$ be variables with alphabets $\mathcal{A}_{A}$,
$\mathcal{A}_{B}$ and $\mathcal{A}_{C}$, respectively. We denote
particular values of $A$, $B$ and $C$ using lower-case letters
$a$, $b$ and $c$. Without loss of generality, we assume that $\mathcal{A}_{A}$,
$\mathcal{A}_{B}$ and $\mathcal{A}_{C}$ are subsets of the natural
numbers (including 0). 

Let $\mathcal{D}$ be defined by the Cartesian product $\mathcal{D}\doteq\mathcal{A}_{A}\times\mathcal{A}_{B}\times\mathcal{A}_{C}$.
Let $f$ be a Boolean function, $f:\mathcal{D}\rightarrow\left\{ 0,1\right\} $,
and define the set $S\doteq\left\{ \left(a,\, b,\, c\right)\in\mathcal{D}\,\,:\,\, f\left(a,\, b,\, c\right)=0\right\} $.
The function $f$ induces mappings $f_{A}$, $f_{B}$ and $f_{C}$,
defined by \[
f_{C}\left(a,\, b\right)\doteq\left\{ c\in\mathcal{A}_{C}\,\,:\,\,\left(a,\, b,\, c\right)\in S\right\} .\]
 The mappings $f_{A}$ and $f_{B}$ are defined similarly.

If $f_{A}$, $f_{B}$ and $f_{C}$ are all functions, then $f$ is
said to be a \emph{constraint function}. For some combination $\left(a,\, b,\, c\right)$,
if $f\left(a,\, b,\, c\right)=0$ then the constraint is said to be
\emph{satisfied}. It is sometimes convenient to refer to the set $S$
as the \emph{satisfaction} of the constraint function $f$. 

A constraint function has a one-to-one correspondence with its satisfaction.
The satisfaction can be written as a table with three columns, which
we call the \emph{satisfaction table}. 

\begin{figure}[hbt]
\begin{center}\begin{tabular}{cc>{\centering}p{1in}}
\begin{minipage}[c]{1.5in}%
\includegraphics[%
  width=1.5in]{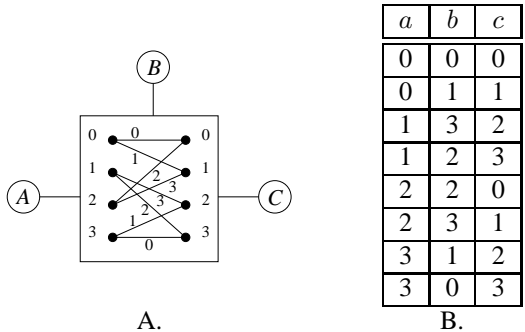}\end{minipage}%
&
&
\begin{minipage}[c]{1in}%
\begin{center}\begin{tabular}{|c|c|c|}
\hline 
$a$&
$b$&
$c$\tabularnewline
\hline
\hline 
0&
0&
0\tabularnewline
\hline 
0&
1&
1\tabularnewline
\hline 
1&
3&
2\tabularnewline
\hline 
1&
2&
3\tabularnewline
\hline 
2&
2&
0\tabularnewline
\hline 
2&
3&
1\tabularnewline
\hline 
3&
1&
2\tabularnewline
\hline 
3&
0&
3\tabularnewline
\hline
\end{tabular}\end{center}\end{minipage}%
\tabularnewline
A.&
&
B.\tabularnewline
\end{tabular}\end{center}

\caption{\label{fig: An example trellis section} An example constraint node
(A), showing a detailed trellis description of its constraint; and
(B) the set $S$ corresponding to this constraint.}
\end{figure}

The satisfaction table corresponds to a \emph{trellis graph}, which
consists of two columns of states and a set of branches connecting
between them. On the left of the trellis, the states correspond to
the alphabet $\mathcal{A}_{A}$. The states on the right correspond
to $\mathcal{A}_{C}$. The branches in the middle are labeled with
symbols from $\mathcal{A}_{B}$. For each row $\left(a,\, b,\, c\right)$
in the satisfaction table, there is a branch in the trellis which
connects $a$ with $c$, and which is labeled $b$. This relationship
is illustrated by Fig. \ref{fig: An example trellis section}.

When we present a constraint graph, we include an unlabeled trellis
graph for each constraint node. This trellis graph indicates the structure
and complexity of the corresponding constraint function. 

\subsection{\label{sub:The-sum-product-update}The sum-product update rule.}

The sum-product algorithm replaces each variable in the constraint
graph with a random variable. The task is then to compute the (conditional)
probability mass of one variable, given the available independent
(conditional) probability masses of all other variables.

Each edge in the graph is then associated with two probability masses,
because each edge is connected to two nodes. Each function node produces,
for every edge, a unique estimate of the (conditional) probability
mass, based on the information which appears at the other edges. Probability
masses are generally represented by vectors, and may sometimes be
represented by summary messages, such as log-likelihood ratios or
soft bits.

Let $\rho_{A}$, $\rho_{B}$ and $\rho_{C}$ be the conditional probability
mass vectors for variables $A$, $B$ and $C$, respectively, and
let $\rho_{A}\left(i\right)$ denote the $i^{\textrm{th}}$ element
of $\rho_{A}$. Define $S_{C=c}$ as the set $S_{C=c}\doteq\left\{ \left(a,\, b\right)\,\,:\,\,\left(a,\, b,\, c\right)\in S\right\} $.
For most applications, the sum-product algorithm is described by the
equation \begin{equation}
\rho_{C}\left(c\right)\propto\sum_{\left(a,b\right)\in S_{C=c}}\rho_{A}\left(a\right)\rho_{B}\left(b\right).\label{eq: sum-product proportionality}\end{equation}

The proportionality symbol in (\ref{eq: sum-product proportionality})
is used to indicate that the resulting mass vector must be normalized
so that the sum is equal to one. Let $S_{C}$ be the union $S_{C}\doteq\bigcup_{c\in\mathcal{A}_{C}}S_{C=c}$.
The normalized sum-product equation is then \begin{eqnarray}
\rho_{C}\left(c\right) & = & \frac{\sum_{\left(a,b\right)\in S_{C=c}}\rho_{A}\left(a\right)\rho_{B}\left(b\right)}{\sum_{\left(a,b\right)\in S_{C}}\rho_{A}\left(a\right)\rho_{B}\left(b\right)}\nonumber \\
 & = & \frac{\sum_{\left(a,b\right)\in S_{C=c}}\rho_{A}\left(a\right)\rho_{B}\left(b\right)}{1-\sum_{\left(a,b\right)\in\overline{S_{C}}}\rho_{A}\left(a\right)\rho_{B}\left(b\right)}.\label{eq: normalized sum-product}\end{eqnarray}

The node update rule (\ref{eq: normalized sum-product}) is used to
update the message transmitted on each edge from each function node.

Throughout this paper, we will assume the use of a {}``flooding''
schedule for message passing. In this schedule, all nodes receive
messages at the start of each iteration. They compute updated messages
on each edge, and then complete the iteration by transmitting the
updated messages.

\subsection{\label{sub:The-relaxation-update}The relaxation update rule.}

In the conventional sum-product algorithm, a node's outgoing messages
are \emph{replaced} by a new probability mass determined according
to (\ref{eq: normalized sum-product}). One alternative to replacement
is the method of successive relaxation \cite{Hemait:DynamicsSOR_ISIT2004},
which we refer to simply as \emph{relaxation}.

Let $v_{C}$ be the mass estimated according to (\ref{eq: normalized sum-product}).
Also let $\rho_{C}$ refer to the mass produced by relaxation at time
$t$, and let $\rho_{C}^{\prime}$ be the mass produced by relaxation
at time $t-1$. Then the relaxation update rule is described by \begin{equation}
\rho_{C}=\rho_{C}^{\prime}+\beta\left(\nu_{C}-\rho_{C}^{\prime}\right).\label{eq: relaxation update rule}\end{equation}
 where $\beta$ is referred to as the \emph{relaxation parameter},
$0<\beta<1$. 

The relaxation method produces smoother updating than the replacement
method. It has been shown that relaxation (\ref{eq: relaxation update rule})
is equivalent to Euler's method for simulating differential equations.
It has also been shown that the relaxation method results in superior
performance for some codes.

\section{\label{sec:The-Stochastic-Algorithm}The Stochastic Algorithm}

The stochastic decoding algorithm is a message-passing algorithm.
Like the sum-product algorithm, it is based on the code's constraint
graph. In the stochastic algorithm, messages are not probability mass
vectors. Stochastic messages are integers which change randomly according
to some probability mass. 

Stochastic messages can be regarded as sequences of integers. The
probability mass is communicated over time by the sequence. One mass
corresponds to many possible sequences.

At each discrete time instant $t$, a function node, $N$, receives
messages $A_{r}\left(t\right)$, $B_{r}\left(t\right)$ and $C_{r}\left(t\right)$,
and transmits messages $A_{t}\left(t\right)$, $B_{t}\left(t\right)$
and $C_{t}\left(t\right)$. The integer messages vary randomly over
time, according to the probability masses $\rho_{A}$, $\rho_{B}$
and $\rho_{C}$, respectively. The message $A_{r}\left(t\right)$
therefore equals integer $a$ with probability $\rho_{A}\left(a\right)$.
All probabilities lie in the open interval $\left(0,\,1\right)$.

\subsection{\label{sub:The-stochastic-update} The stochastic update rule.}

To implement the stochastic message-passing algorithm, we use the
following deterministic message update rule at each function node.
We consider the propagation of message from $\left(A_{r}\left(t\right),\, B_{r}\left(t\right)\right)$
to $C_{t}\left(t\right)$. Suppose that, at time $t$, $A_{r}\left(t\right)=a$
and $B_{r}\left(t\right)=b$. Then \begin{equation}
C_{t}\left(t\right)=\left\{ \begin{array}{ll}
f_{C}\left(a,\, b\right) & \textrm{if \,}\left(a,\, b\right)\in S\\
C_{t}\left(t-1\right) & \textrm{otherwise.}\end{array}\right.\label{eq: stochastic update rule}\end{equation}

We claim that the stochastic update rule (\ref{eq: stochastic update rule})
produces a message $C$ with mass $\rho_{C}$ which is equal to the
mass computed by the sum-product update rule (\ref{eq: normalized sum-product}).
To support this claim, we rely upon the following assumptions:

\begin{enumerate}
\item There is no correlation between any pair of sequences received by
a constraint node.
\item The probability mass associated with any message sequence is \emph{stationary}. 
\end{enumerate}
When a node is considered in isolation from the graph, the above conditions
apply.

\begin{thm}
\label{thm: sum-product equals stochastic} A node $N$ receives stochastic
messages $A_{r}\left(t\right)$ and $B_{r}\left(t\right)$, and transmits
a message $C_{t}\left(t\right)$ according to the update rule (\ref{eq: stochastic update rule}).
The received messages have stationary probability masses $\rho_{A}$
and $\rho_{B}$. Then the mass of the transmitted message, $\rho_{C}$,
is equal to the result computed by the sum-product update rule.
\end{thm}
\begin{proof}
The probability of transmitting a particular integer $c$ is \begin{eqnarray*}
\rho_{C}\left(c\right) & = & \sum_{\left(a,b\right)\in S_{C=c}}\rho_{A}\left(a\right)\rho_{B}\left(b\right)\\
 &  & +\left[\sum_{\left(a,b\right)\in\overline{S_{C}}}\rho_{A}\left(a\right)\rho_{B}\left(b\right)\right]\textrm{Pr}\left(C_{t}\left(t-1\right)=c\right).\end{eqnarray*}
 We assume that the mass of each variable is stationary. Then $\textrm{Pr}\left(C_{t}\left(t-1\right)=c\right)=\rho_{C}\left(c\right)$.
Therefore \[
\rho_{C}\left(c\right)=\frac{\sum_{\left(a,b\right)\in S_{C=c}}\rho_{A}\left(a\right)\rho_{B}\left(b\right)}{1-\sum_{\left(a,b\right)\in\overline{S_{C}}}\rho_{A}\left(a\right)\rho_{B}\left(b\right)},\]
 which is the same as (\ref{eq: normalized sum-product}).
\end{proof}
The stochastic update rule (\ref{eq: stochastic update rule}) results
in very simple node implementations. The implementation of a node
consists of a logic gate which is precisely as complex as the node's
trellis description.

\subsection{\label{sub:Message-passing-implementation-on}Message-passing implementation
on acyclic graphs.}

For a constraint graph with no cycles, a complete stochastic decoder
is constructed by applying the update rule (\ref{eq: stochastic update rule})
to every constraint node. Probability masses, representing the channel
information, are stored at the variable nodes. The variable nodes
generate the first messages randomly, according to the channel information. 

The random integer messages flow deterministically through the decoder,
through a cascade of constraint nodes, until they arrive back at the
variable nodes. The messages are tabulated in histograms as they arrive
at the variable nodes. After $l$ time-steps, the variable nodes make
decisions by selecting the symbol with the largest count.

When implemented in this way, the assumptions of Sec. \ref{sub:The-stochastic-update}
are never violated. The probability mass of each message in the decoder
is therefore equivalent to that of the sum-product algorithm. By increasing
$l$, the result of stochastic decoding can be made arbitrarily close
to that of sum-product decoding.

\begin{example}
\label{exa:An-acyclic-constraint} An acyclic constraint graph for
a (16, 11) Hamming code is shown in Fig. \ref{fig: (16,11) graph}.
A stochastic decoder is constructed by replacing each constraint node
with logic gates, according to the update rule (\ref{eq: stochastic update rule}).
We allow $l=250$ time-steps for decoding. The performance of the
resulting code for antipodal transmission on a Gaussian channel is
shown in Fig. \ref{fig: (16,11) decoder results}.

\begin{figure}[hbt]
\begin{center}\includegraphics[%
  width=3.2in]{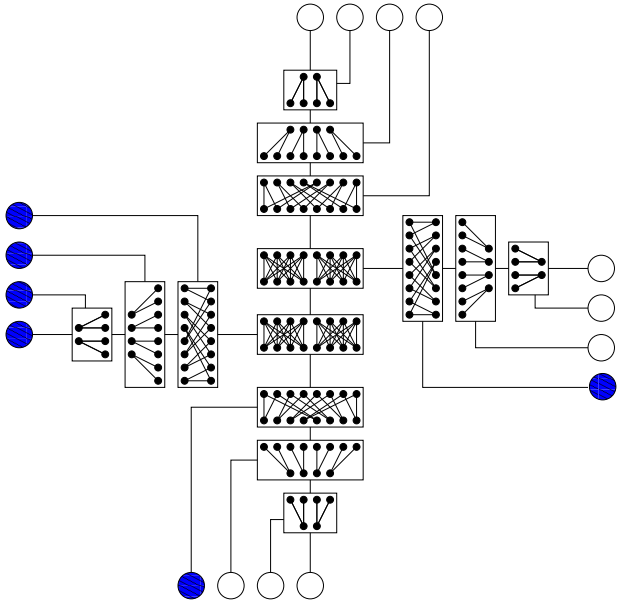}\end{center}

\caption{\label{fig: (16,11) graph} An acyclic constraint graph for the (16,
11) Hamming code. Dark circles represent parity bits. Light circles
represent information bits.}
\end{figure}

\begin{figure}[hbt]
\begin{center}\includegraphics[%
  width=0.85\columnwidth]{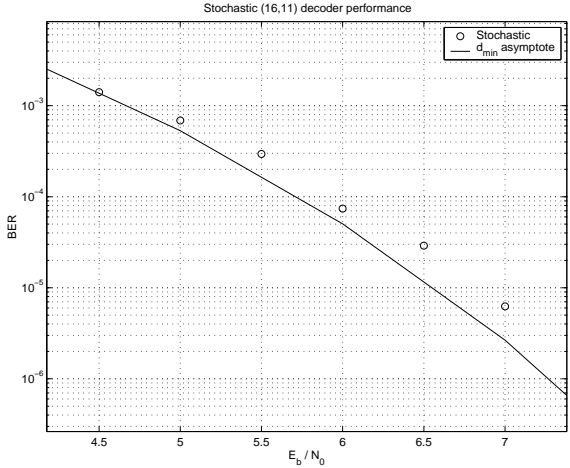}\end{center}

\caption{\label{fig: (16,11) decoder results} Performance results for a (16,11)
Hamming stochastic decoder, based on the graph in Fig. \ref{fig: (16,11) graph}.
Also shown is the code's minimum-distance asymptote. Each data point
represents 50 errors.}
\end{figure}

\end{example}

\subsection{\label{sub:Graphs-with-cycles}Graphs with cycles and Supernodes.}

If the constraint graph contains one or more cycles, then the assumptions
of Sec. \ref{sub:The-stochastic-update} are violated. In particular,
the independence among received messages $A_{r}\left(t\right)$, $B_{r}\left(t\right)$
and $C_{r}\left(t\right)$ no longer holds for some or all constraint
nodes. 

When the messages at a node are correlated, it is possible for a group
of nodes to settle into a fixed state. This state is maintained solely
by the messages within the cycle. No pattern of independent messages
is sufficient to restore the cycle to proper functioning. We refer
to this phenomenon as \emph{latching}. 

A simple latching example is the all-zero state in an LDPC graph.
If all internal messages between parity and equality nodes are zero,
then they will be held permanently at zero, regardless of any activity
from the variable nodes.

To avoid latching, cycles may be interrupted with a special stochastic
node, called a \emph{supernode}. A supernode receives stochastic messages
and tabulates them in histograms. It then generates new messages based
on the estimated probability masses, resulting in a new uncorrelated
sequence.

\begin{figure}[hbt]
\begin{center}\includegraphics[%
  width=0.70\columnwidth]{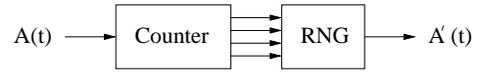}\end{center}

\caption{\label{fig: supernode} A simple supernode implementation. A stochastic
sequence $A\left(t\right)$ is converted into an uncorrelated sequence
$A'\left(t\right)$ which encodes the same mass. `RNG' denotes a random
number generator, which generates random numbers according to the
mass estimated for $A\left(t\right)$.}
\end{figure}

A supernode can be implemented with counters and a linear feedback
shift-register (LFSR) to generate random numbers. The simplest supernode
implementation is illustrated in Fig. \ref{fig: supernode}.

We also claim that the supernode must be \emph{packetized}, which
is defined as follows. Suppose a supernode is generating a random
sequence according to an estimated mass $\rho$. We say that the supernode's
behavior is packetized if the estimated mass $\rho$ is only updated
every $l$ time-steps. During these $l$ time-steps, the supernode
tabulates the received message(s). After the $l^{\textrm{th}}$ time-step,
$\rho$ is updated with the newly tabulated information. 

Data received during the $l$ time-steps is called a \emph{packet},
and the $l$-step time interval is called an \emph{iteration}. There
are two obvious procedures by which $\rho$ may be updated:

\begin{enumerate}
\item The tabulation is cleared after each iteration, and the mass estimate
is \emph{replaced} by a new estimate representing only the data from
the most recent packet.
\item The tabulation is not cleared after each packet of $l$ time-steps.
The tabulation is \emph{accumulated} over many iterations until decoding
is complete.
\end{enumerate}
For sufficiently large $l$, these supernode update rules approach
the behavior of sum-product and relaxation updates, respectively.

\begin{thm}
Suppose each cycle in a stochastic constraint graph contains a supernode,
and the mass estimated by the supernode is updated by \emph{replacement}
every $l$ time-steps. Then the result of stochastic decoding approaches
that of sum-product decoding as $l$ is increased.
\end{thm}
\begin{proof}
Let $l$ be the total number of observations of an event $C$, and
let $T_{C}\left(c\right)$ be the total number of observations for
which $C=c$. By definition for a stationary mass $\rho$, $\lim_{l\rightarrow\infty}T_{C}\left(c\right)/l=\rho\left(c\right)$.
The mass of $C$ is is therefore estimated with arbitrarily small
error as $l$ is increased. 

The messages produced by the supernode have a fixed mass, and are
therefore stationary. They are also not correlated with any other
messages in the graph. Theorem \ref{thm: sum-product equals stochastic}
therefore applies, and the stochastic decoder approaches the behavior
of a sum-product iteration for large $l$.
\end{proof}
\begin{thm}
Suppose each cycle in a stochastic constraint graph contains a supernode,
and the mass estimated by the supernode is updated by \emph{accumulation}
every $l$ time-steps. Then the result of stochastic decoding approaches
that of relaxation, where the relaxation parameter $\beta$ varies
as $1/m$, where $m$ is the number of iterations.
\end{thm}
\begin{proof}
Let $\rho_{C}$ be a supernode's newly updated estimate of the mass
of an event $C$ just after $m$ iterations, and let $\rho_{C}^{\prime}$
be the supernode's estimate after $m-1$ iterations. The supernode
has made $ml$ total observations. Let $N_{c}$ be number of observations
in the $m^{\textrm{th}}$ packet for which $C=c$. Let $N_{c}^{\prime}$
be the number of observations in the first $m-1$ packets for which
$C=c$. Then the new estimate is \begin{eqnarray}
\rho_{C} & = & \frac{N_{c}+N_{c}^{\prime}}{ml}\nonumber \\
 & = & \frac{1}{m}\left(\frac{N_{c}}{l}\right)+\frac{m-1}{m}\left(\frac{N_{c}^{\prime}}{\left(m-1\right)l}\right)\nonumber \\
 & = & \rho_{C}^{\prime}+\frac{1}{m}\left(\frac{N_{c}}{l}-\rho_{C}^{\prime}\right).\label{eq: stochastic SOR}\end{eqnarray}
 As $l\rightarrow\infty$, $N_{c}/l$ approaches the sum-product estimate
of $\rho_{c}$ for the $m^{\textrm{th}}$ iteration. Upon substitution,
we find that (\ref{eq: stochastic SOR}) approaches the relaxation
update rule with $\beta=1/m$. The difference between relaxation and
(\ref{eq: stochastic SOR}) can be made arbitrarily small by increasing
$l$.
\end{proof}
\begin{example}
The (16,11) Hamming code graph of Ex. \ref{exa:An-acyclic-constraint}
can be used to construct a (256,121) Block Turbo Code. The decoder
for this code consists of 32 Hamming decoders, arranged in 16 rows
and 16 columns. Each row shares precisely one bit with each column.
Where two decoders share a bit, they are joined by an \emph{equality}
constraint. 

\begin{figure}[hbt]
\begin{center}\includegraphics[%
  width=0.70\columnwidth]{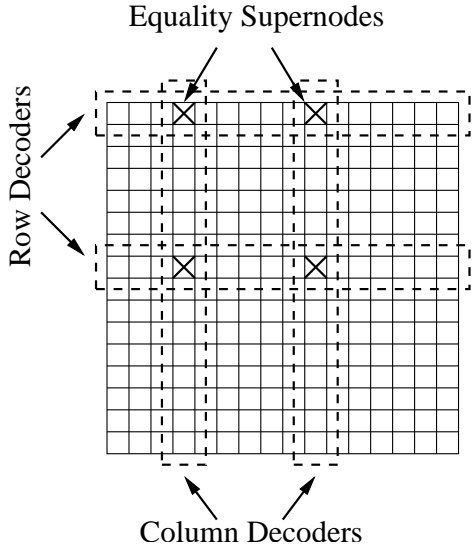}\end{center}

\caption{\label{fig: product codeword} The structure of a product code. Each
row represents an instance of the decoder from Fig. \ref{fig: (16,11) graph}.
Each column represents another instance of the decoder. Where a row
and a column cross, they are joined by an equality supernode. The
code's variable nodes also connect to the equality supernode.}
\end{figure}

The equality constraints create cycles in the constructed graph. To
resolve this, we implement each equality node as a supernode. Within
the equality supernode, received messages are converted to probability
mass estimates. Using these mass estimates, the outgoing masses are
computed using the conventional sum-product update rule (\ref{eq: normalized sum-product}).
New stochastic sequences are generated to conform to the outgoing
mass estimate.

\begin{figure}[hbt]
\begin{center}\includegraphics[%
  width=0.65\columnwidth]{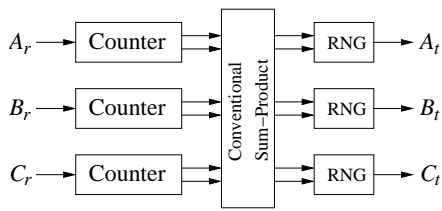}\end{center}

\caption{\label{fig: equality supernode} Structure of the equality supernode.
The counters update by accumulation. The conventional sum-product
calculation is invoked every 250 time-steps.}
\end{figure}

We choose a flooding schedule with an \emph{accumulation} update rule,
and a packet length of $l=250$. Eight iterations are allowed, for
a total decoding time of 2000 time-steps. The stochastic decoder's
performance is as shown in Fig. \ref{fig: product decoder performance}.
Also shown is the simulated performance of a commercial Block Turbo
decoder from Comtech/Advanced Hardware Architectures \cite{AHA}.
The AHA decoder uses 26 iterations for decoding, and six bits of precision
for the channel samples.

\begin{figure}[hbt]
\begin{center}\includegraphics[%
  width=0.85\columnwidth]{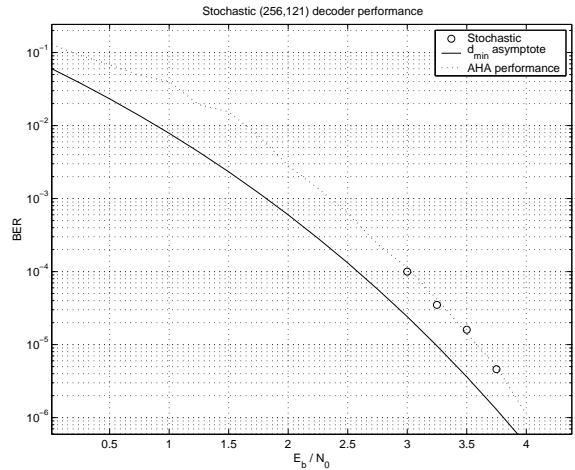}\end{center}

\caption{\label{fig: product decoder performance} Performance of a stochastic
(256,121) Block Turbo decoder. Also shown is simulated data for a
similar Block Turbo decoder produced by Comtech AHA. Each data point
represents 50 errors.}
\end{figure}

\end{example}

\section{\label{sec:Conclusions}Discussion and Conclusions}

While this paper validates that stochastic iterative decoding can
be applied to interesting codes (e.g., Block Turbo codes), it also
raises an interesting set of ancillary questions. There are numerous
alternative stochastic update rules and message schedules, and many
possible simplifications to the algorithm. Questions also remain concerning
the relationships among the packet size, the number of iterations
and the performance of a stochastic decoder for a given graph. 

These issues hint at a broad set of interesting research questions,
with significant potential to improve the complexity of iterative
decoders.

\bibliographystyle{IEEEbib}
\bibliography{winstead-isit-submission-2005}

\end{document}